\documentclass[a4paper]{article}
\usepackage{amssymb}
\usepackage{latexsym}
\usepackage{amsmath}
\usepackage{graphicx}

\usepackage{esdiff}
\usepackage{mathtools}

\def\vp{\varphi}
\def\ve{\varepsilon}
\def\al{\alpha}
\def\nb{\nabla}
\def\th{\theta}
\def\e{\eta}
\def\l{\lambda}
\def\k{\kappa}
\def\L{\Lambda}
\def\b{\beta}
\begin{document}
\title{Black hole solutions in gravity with nonminimal derivative coupling and nonlinear material fields}
\author{Mykola M. Stetsko \\{\small E-mail: mstetsko@gmail.com}\\{\small Department for Theoretical Physics, Ivan Franko National University of Lviv,}\\{\small 12 Drahomanov Str., Lviv, UA-79005, Ukraine}}

\maketitle
\abstract{Scalar-tensor theory of gravity with nonlinear electromagnetic field, minimally coupled to gravity is considered and static black hole solutions are obtained. Namely, power-law and Born-Infeld nonlinear Lagrangians for  the electromagnetic field are examined. Since the cosmological constant is taken into account, it allowed us to investigate so-called topological black holes. Black hole thermodynamics is studied, in particular temperature of the black holes is calculated and examined and the first law of thermodynamics is obtained with help of Wald's approach.}

\section{Introduction}
General Relativity is very successful theory with solid experimental grounds which confirm the theory with extremely high precision. It explains  physical phenomena on very wide space and time scales from planetary motion up to the evolution of the Universe \cite{Will_LRR2014, Berti_CQG2015}. In spite of its there are some open questions that still remain unsolved. Among the most important open issues we distinguish dark energy/dark matter problem, origin of curvature singularities, description of inflationary epoch which do not have consistent solution in the framework of standard General Relativity and possibly open some windows (or bring some hints) for development more general theory of gravity \cite{Clifton_PhysRept2012, Heisenberg_PhRept2019}. We point out here that there are plenty of approaches to solve the mentioned puzzles, some of them suggest radical revision of basic principles on which theory of gravity should be grounded, whereas other approaches are more down-to-earth and consider just minimal modification of mainly successful General Relativity.

Scalar-tensor theories of gravity belong to the latter class of approaches where in addition to standard gravitational degrees of freedom some new degrees of freedom are taken into account which are represented by scalar fields. Among the different types of scalar-tensor theories so-called Horndeski gravity is worth of special attention. Proposed several decades ago \cite{Horndeski_IJTP74}  Horndeski gravity has received considerable attention only in recent decade, when string theory inspired models gave rise to the formulation of Galileon model \cite{Deffayet_PRD09}. We point out here that one of the most attractive features of Horndeski gravity is the absence of ghosts due to the second order equations of the theory.

It should be noted that the general structure of Horndeski gravity is quite complicated and it is  supposed to be the main obstacle that  prevent applying it in its most general setup for solution and analysis of some particular problems in cosmology and General Relativity. Thus some particular cases of general Horndeski Gravity are usually used for this aim. Namely, the theories with nonminimal derivative coupling between gravitational and scalar degrees of freedom has attracted considerable attention for recent years. This model was used for broad class of problems in Cosmology and Physics of black holes. In particular, new cosmological solutions were found \cite{Sushkov_PRD09}, various aspects of inflation were studied \cite{Saridakis_PRD10, Germani_PRL10}, new black holes' solutions and many features of Black Hole Physics were also examined \cite{Rinaldi_PRD12, Minamitsuji_PRD14,Feng_JHEP15,Stetsko_arx18,Stetsko_PRD19}.

Our work is devoted to black holes in the theory with nonminimal derivative coupling with additional material field which are represented by nonlinear Lagrangians, namely we consider two types of nonlinear field, one of them is so-called power-law case and the other one is of Born-Infeld type. We derive  static black holes solutions, investigate their main features and we also pay some attention to the thermodynamics of the obtained black holes. We also point out that more about power-law case can be found in a recently published paper \cite{Stetsko_PRD19}.

\section{Field equations and their black hole solutions}
The starting point in our work is an action integral which consists of several terms, namely it contains standard Einstein-Hilbert term with cosmological constant, the second part corresponds to the scalar field which consists of two terms with minimal and nonminimal coupling to gravity and finally there is a contribution of the mentioned gauge field (electromagnetic) which is minimally coupled to gravity. As a consequence the action can be written in the form:
\begin{equation}\label{action}
S=\int d^{n+1}x\sqrt{-g}\left( R-2\Lambda-\frac{1}{2}\left(\alpha g^{\mu\nu}-\eta G^{\mu\nu}\right)\partial_{\mu}\vp\partial_{\nu}\vp +{\cal L}_m\right),
\end{equation}
where $g_{\mu\nu}$  and $g=det{g_{\mu\nu}}$ are the metric tensor and its determinant respectively, $G_{\mu\nu}$ and $R$ are the Einstein tensor and Ricci scalar correspondingly, $\L$ denotes the cosmological constant, $\vp$ is the scalar field, nonminimally coupled to gravity and finally ${\cal L}_m$ is the Lagrangian minimally coupled to gravity. As we mentioned in the introduction we examine two types of nonlinear electromagnetic field, and the material Lagrangians can be written in the form:
\begin{eqnarray}\label{matter_lagran}
{\cal L}_m=
\begin{cases}
(-F_{\mu\nu}F^{\mu\nu})^p, \quad PMI{} \\\\
4\b^2\left(1-\sqrt{1+\frac{F_{\mu\nu}F^{\mu\nu}}{2\b^2}}\right),  \quad BI{} 
\end{cases}
\end{eqnarray}
and here the upper Lagrangian corresponds to the so-called power-law nonlinearity, whereas the lower one represents well-known Born-Infeld Lagrangian. 

Equations of motion can be easily derived from the action (\ref{action}) with account of the matter Lagrangians (\ref{matter_lagran}). So we write:
\begin{equation}\label{eom}
G_{\mu\nu}+\Lambda g_{\mu\nu}=\frac{1}{2}(\alpha T^{(1)}_{\mu\nu}+\eta T^{(2)}_{\mu\nu})+T^{(3)}_{\mu\nu},
\end{equation}
and here we use the following notations
\begin{equation}\label{scal_min}
T^{(1)}_{\mu\nu}=\nb_{\mu}\vp\nb_{\nu}\vp-\frac{1}{2}g_{\mu\nu}\nb^{\lambda}\vp\nb_{\lambda}\vp,
\end{equation}
\begin{eqnarray}\label{scal_nm}
\nonumber T^{(2)}_{\mu\nu}=\frac{1}{2}\nb_{\mu}\vp\nb_{\nu}\vp R-2\nb^{\lambda}\vp\nb_{\nu}\vp R_{\lambda\mu}+\frac{1}{2}\nb^{\lambda}\vp\nb_{\lambda}\vp G_{\mu\nu}-g_{\mu\nu}\left(-\frac{1}{2}\nb_{\lambda}\nb_{\kappa}\vp\nb^{\lambda}\nb^{\kappa}\vp\right.\\\left.+\frac{1}{2}(\nb^2\vp)^2-R_{\lambda\kappa}\nb^{\lambda}\vp\nb^{\kappa}\vp\right)
-\nb_{\mu}\nb^{\lambda}\vp\nb_{\nu}\nb_{\lambda}\vp+
\nb_{\mu}\nb_{\nu}\vp\nb^2\vp-R_{\lambda\mu\kappa\nu}\nb^{\lambda}\vp\nb^{\kappa}\vp,
\end{eqnarray}
and $T^{(3)}_{\mu\nu}$ denotes stress-energy tensor for electromagnetic field, the evident form of which depends on the type of Lagrangian we consider. Namely, for power-law case we can write:
\begin{equation}\label{max_tr_nlin}
T^{(3)}_{\mu\nu}=\frac{g_{\mu\nu}}{2}\left(-F_{\l\k}F^{\l\k}\right)^p+2p\left(-F_{\l\k}F^{\l\k}\right)^{p-1}F_{\mu\rho}{F_{\nu}}^{\rho},
\end{equation}
whereas for Born-Infeld type we have:
\begin{equation}\label{max_tr_nlin_BI}
T^{(3)}_{\mu\nu}=2\b^2g_{\mu\nu}\left(1-\sqrt{1+\frac{F_{\k\l}F^{\k\l}}{2\b^2}}\right)+\frac{2F_{\mu\rho}{F_{\nu}}^{\rho}}{\sqrt{1+\frac{F_{\k\l}F^{\k\l}}{2\b^2}}}.
\end{equation} 
Equations of motion for the scalar and electromagnetic fields can be written as follows:
\begin{equation}\label{scal_f_eq}
(\alpha g_{\mu\nu}-\eta G_{\mu\nu})\nb^{\mu}\nb^{\nu}\vp=0.
\end{equation}
\begin{equation}\label{Maxwell_eq}
\nb_{\mu}\left({-\cal F}^{p-1}F^{\mu\nu})\right)=0 \quad (PMI), \quad \nb_{\mu}\left(\frac{F^{\mu\nu}}{\sqrt{1+{\cal F}/2\b^2}}\right)=0 \quad (BI),
\end{equation}
where ${\cal F}=F_{\mu\nu}F^{\mu\nu}$.

We are going to obtain static solutions and corresponding metrics are supposed to take the form:
\begin{equation}\label{metric}
ds^2=-U(r)dt^2+W(r)dr^2+r^2d\Omega^{2(\ve)}_{(n-1)},
\end{equation}
where $d\Omega^{2(\ve)}_{n-1}$ is supposed to be written as follows: 
\begin{eqnarray}
d\Omega^{2(\ve)}_{(n-1)}=
\begin{cases}
d\th^2+\sin^2{\th}d\Omega^2_{(n-2)}, \quad \ve=1,\\
d\th^2+{\th}^2 d\Omega^2_{(n-2)},\quad \ve=0,\\
d\th^2+\sinh^2{\th}d\Omega^2_{(n-2)},\quad \ve=-1,
\end{cases}
\end{eqnarray}
and here $d\Omega^2_{(n-2)}$ is the line element of a $n-2$ -- dimensional sphere. We point out here that since we take into account the cosmological constant it is possible to examine so-called topological black holes also, where the horizon surface might be a surface of negative or zeroth curvature.

We investigate only electrically charged black holes, so we suppose that the gauge potential takes simple form: $A=A_0(r)dt$ Having chosen the form of the metric (\ref{metric}) and equations for electromagnetic field (\ref{Maxwell_eq}) one can easily integrate them and as a result we can write:
\begin{equation}\label{EM_fields}
F_{rt}=\frac{q}{r^{(n-1)/(2p-1)}}\sqrt{UW} \quad (PMI),\quad
F_{rt}=\frac{q\b}{\sqrt{q^2+\b^2r^{2(n-1)}}}\sqrt{UW} \quad (BI).
\end{equation}
The  equation for the scalar field (\ref{scal_f_eq}) can be easily integrated just for a once and for simplicity we assume that $\al g_{rr}-\e G_{rr}=0$, for more detailed reasoning one might consult for example \cite{Minamitsuji_PRD14,Stetsko_arx18,Stetsko_PRD19}.

\begin{figure}
\centerline{\includegraphics[scale=0.3,clip]{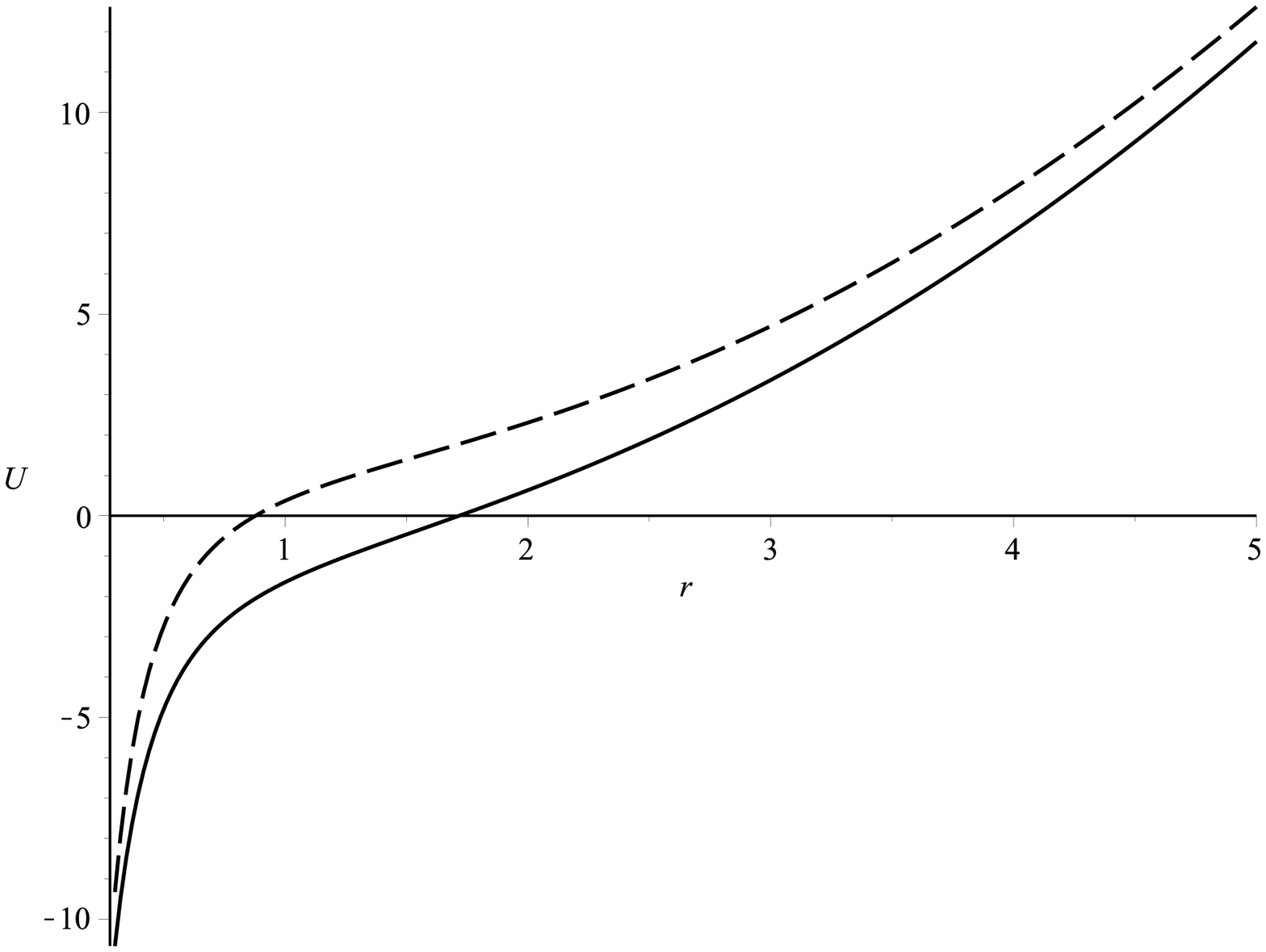}\includegraphics[scale=0.29,clip]{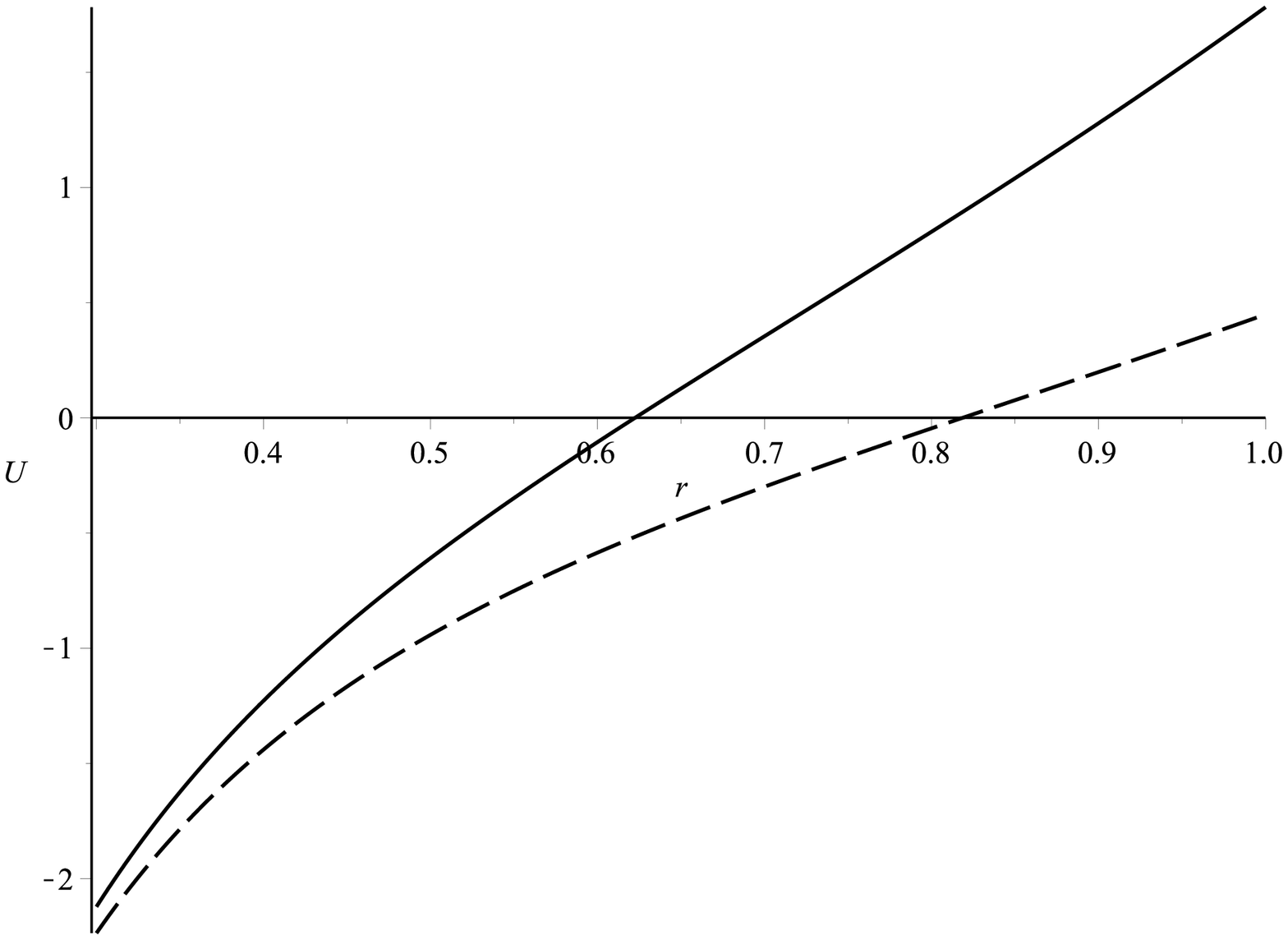}}
\caption{Metric functions $U(r)$ for power-law (the left graph) and Born-Infeld field (the right graph).}\label{metr_f_graph}%On the left graph the solid and dashed curves correspond to $p=1$ and $p=(n+1)/4$ respectively. For the left graph all the fixed parameters are taken as follows: $n=4$, $\al=0.2$, $\e=0.4$, $\ve=1$, $\L=-2$, $q=0.2$ and $\mu=1$, solid curve corresponds to $p=1$ and the dashed one to $p=(n+1)/4$. For the Born-Infeld case (the right graph) the fixed parameters takes the values: $n=3$, $\al=0.2$, $\e=0.4$, $\ve=1$, $\L=-4$, $q=0.1$, $m=1$, and here solid curve corresponds to $\b=1$, whereas the dashed curve represents limit case of linear field $\b\rightarrow\infty$.
\end{figure}

Now, using the evident form of the electromagnetic fields and the condition imposed on the $G_{rr}$ component, we obtain the relations for the square of derivative $\vp'$, product of the metric functions $UW$ and finally write a relation for the metric function $U(r)$.  For the power-law case we can obtain:
\begin{equation}\label{vp_pmi}
(\vp\rq{})^2=-\frac{4r^2W}{\e(2\al r^2+\ve\e(n-1)(n-2))}\left(\al+\L\e+\frac{2^{p-1}(2p-1)\e q^{2p}}{r^{2p(n-1)/(2p-1)}}\right),
\end{equation}
\begin{equation}\label{UW_pr_pmi}
UW=\frac{\left((\al-\L\e)r^2+\ve\e(n-1)(n-2)-2^{p-1}\e(2p-1)q^{2p}r^{2(1-\frac{p(n-1)}{2p-1})}\right)^2}{(2\al r^2+\ve\e(n-1)(n-2))^2},
\end{equation}
and finally the metric function can be represented in the form:
 \begin{eqnarray}\label{U_int_PMI}
\nonumber U(r)=\ve-\frac{\mu}{r^{n-2}}-\frac{2\L}{n(n-1)}r^{2}-2^p\frac{(2p-1)^2q^{2p}}{(n-1)(2p-n)}r^{2\left(1-\frac{p(n-1)}{2p-1}\right)}+\\\nonumber\frac{(\al+\L\e)^2}{2\al\e(n-1)r^{n-2}}\int\frac{r^{n+1}}{r^2+d^2}dr+2^{p-1}\frac{(2p-1)(\al+\L\e)q^{2p}}{\al(n-1)r^{n-2}}\times\\
\int\frac{r^{n+1-\frac{2p(n-1)}{2p-1}}}{r^2+d^2}dr+2^{2p-3}\frac{(2p-1)^2\e q^{4p}}{\al(n-1)r^{n-2}}\int\frac{r^{n+1-\frac{4p(n-1)}{2p-1}}}{r^2+d^2}dr
 \end{eqnarray}
 and here $d^2=\frac{\ve\e(n-1)(n-2)}{2\al}$
For Born-Infeld case we obtain:
\begin{eqnarray}\label{vp_der_BI}
(\vp')^2=-\frac{4r^2W\left(\al+\L\e-2\b^2\e+\frac{2\b\e}{r^{n-1}}\sqrt{q^2+\b^2r^{2(n-1)}}\right)}{\e(2\al r^2+\ve\e(n-1)(n-2))};
\end{eqnarray}
\begin{equation}\label{UW_prod_BI}
UW=\frac{\left((\al-\L\e+2\b^2\e)r^2+\ve\e(n-1)(n-2)-\frac{2\b\e} {r^{n-3}}\sqrt{q^2+\b^2r^{2(n-1)}}\right)^2}{(2\al r^2+\ve\e(n-1)(n-2))^2};
\end{equation}
\begin{eqnarray}\label{U_int_BI}
\nonumber U(r)=\ve-\frac{\mu}{r^{n-2}}-\frac{2(\L-2\b^2)}{n(n-1)}r^2-\frac{2\b(\al-\L\e+2\b^2\e)}{\al(n-1)r^{n-2}}\times\\\nonumber\int\sqrt{q^2+\b^2r^{2(n-1)}}dr+\frac{(\al+\L\e-2\b^2\e)^2+4\b^4\e^2}{2\al\e(n-1)\al\e r^{n-2}}\int\frac{r^{n+1}dr}{r^2+d^2}-\\\frac{2\b(\al+\L\e-2\b^2\e)d^2}{\al(n-1)r^{n-2}}\int\frac{\sqrt{q^2+\b^2r^{2(n-1)}}dr}{r^2+d^2}+\frac{2\b^2\e q^2}{\al(n-1)r^{n-2}}\int\frac{r^{3-n}dr}{r^2+d^2}
\end{eqnarray} 
It should be pointed out here that the integrals in the relations (\ref{U_int_PMI}) and (\ref{U_int_BI}) can be written in explicit form, but some of them can be expressed with help of hypergeometric functions $_{2}F_1(a,b;c;z)$ only, in addition the final form might depend on the parity of dimension of the space.  We will give here the evident form of the integrals only for Born-Infeld case, since for power-law case they are given in \cite{Stetsko_PRD19}. Namely, if $n$ is odd, we write:
 \begin{eqnarray}\label{funct_odd}
\nonumber U(r)=\ve-\frac{\mu}{r^{n-2}}-\frac{2(\L-2\b^2)}{n(n-1)}r^2-\frac{2\b^2(\al-\L\e+2\b^2\e)}{\al n(n-1)}r^2{_{2}F_{1}}\left(-\frac{1}{2},\frac{n}{2(1-n)};\frac{2-n}{2(1-n)};-\frac{q^2}{\b^2}r^{2(1-n)}\right)\\+\frac{(\al+\L\e-2\b^2\e)^2+4\b^4\e^2}{2\al\e(n-1)}\left[\sum^{\frac{n-1}{2}}_{j=0}(-1)^j\frac{d^{2j}r^{2(1-j)}}{n-2j}+\frac{(-1)^{\frac{n+1}{2}}d^n}{r^{n-2}}\arctan{\left(\frac{r}{d}\right)}\right]+U^{(o)}_{BI}(r),
\end{eqnarray}
and here the function $U^{(o)}_{BI}(r)$ takes the following form:
\begin{eqnarray}
\nonumber U^{(o)}_{BI}(r)=\frac{2\b^2\e q^2}{\al(n-1)}\left[\sum^{\frac{n-5}{2}}_{j=0}\frac{(-1)^jr^{6-2n+2j}}{(4-n+2j)d^{2(j+1)}}+\frac{(-1)^{\frac{n-3}{2}}}{d^{n-2}r^{n-2}}\arctan\left(\frac{r}{d}\right)\right]-\frac{2\b^2(\al+\L\e-2\b^2\e)d^2}{\al(n-1)}\\\times\sum^{+\infty}_{j=0}\frac{(-1)^j}{n-2(j+1)}\left(\frac{d}{r}\right)^{2j}{_{2}F_{1}}\left(-\frac{1}{2},\frac{n-2(j+1)}{2(1-n)};-\frac{n+2j}{2(1-n)};-\frac{q^2}{\b^2}r^{2(1-n)}\right);
\end{eqnarray} 
whereas for even $n$ we obtain:
and for even $n$ we obtain:
\begin{eqnarray}\label{funct_even_BI}
\nonumber U(r)=\ve-\frac{\mu}{r^{n-2}}-\frac{2(\L-2\b^2)}{n(n-1)}r^2-\frac{2\b^2(\al-\L\e+2\b^2\e)}{\al n(n-1)}r^2{_{2}F_{1}}\left(-\frac{1}{2},\frac{n}{2(1-n)};\frac{2-n}{2(1-n)};-\frac{q^2}{\b^2}r^{2(1-n)}\right)\\+\frac{(\al+\L\e-2\b^2\e)^2+4\b^4\e^2}{2\al\e(n-1)}\left[\sum^{(n-2)/2}_{j=0}(-1)^j\frac{d^{2j}r^{2(1-j)}}{n-2j}+\frac{(-1)^{\frac{n}{2}}d^n}{2r^{n-2}}\ln{\left(\frac{r^2}{d^2}+1\right)}\right]+U^{(e)}_{BI}(r),
\end{eqnarray}
and here 
\begin{eqnarray}
\nonumber U^{(e)}_{BI}(r)=\frac{2\b^2\e q^2}{\al(n-1)}\left[\sum^{\frac{n-6}{2}}_{j=0}\frac{(-1)^jr^{6-2n+2j}}{(4-n+2j)d^{2(j+1)}}+\frac{(-1)^{\frac{n-2}{2}}}{2d^{n-2}r^{n-2}}\ln\left(1+\frac{d^2}{r^2}\right)\right]-\frac{2\b^2(\al+\L\e-2\b^2\e)d^2}{\al(n-1)}\times\\\nonumber\left[\sum^{+\infty}_{j=0}\frac{(-1)^j}{n-2(j+1)}\left(\frac{d}{r}\right)^{2j}{_{2}F_{1}}\left(-\frac{1}{2},\frac{n-2(j+1)}{2(1-n)};-\frac{n+2j}{2(1-n)};-\frac{q^2}{\b^2}r^{2(1-n)}\right)+(-1)^{\frac{n}{2}}\frac{d^{n-2}}{r^{n-2}}\times\right.\\\left.\left(\sum^{+\infty}_{j=1}\frac{(-1)^j}{j!}\left(-\frac{1}{2}\right)_{j}\left(\frac{q}{\b}\right)^{2j}\frac{r^{2(1-n)j}}{2(n-1)j}-\ln{\left(\frac{r}{d}\right)}\right)\right].
\end{eqnarray}
It is worth being noted here that the evident form for the hypergeometric functions in the latter four relations is valid when $r>d$ and $r^{2(n-1)}>q^2/\b^2$, in other cases other forms of hypergeometric functions should be used. The behaviour of the metric functions $U(r)$ is demonstrated on the Fig.[\ref{metr_f_graph}]. Both graphs share some common features, namely for large distances AdS-term for any case is dominating and contribution from the electromagnetic fields becomes negligibly small, so for this case ($r\rightarrow\infty$) the metric function behaves as follows:
\begin{equation}\label{metr_inf_any}
U\sim \frac{(\al-\L\e)^2}{2\al\e n(n-1)}r^2.
\end{equation} 
 In contrast, for very small values of $r$ the behaviour of the metric function depends on the types of material field, horizon geometry and dimension, but in all the cases it is singular.
 
 To characterize singularities of the metric we use Kretschmann scalar which can be written in the form:
\begin{eqnarray}\label{Kr_scalar}
\nonumber R_{\mu\nu\k\l}R^{\mu\nu\k\l}=\frac{1}{UW}\left(\frac{d}{dr}\left[\frac{U'}{\sqrt{UW}}\right]\right)^2+\frac{(n-1)}{r^2W^2}\times\\\left(\frac{(U')^2}{U^2}+\frac{(W')^2}{W^2}\right)+\frac{2(n-1)(n-2)}{r^4W^2}(\ve W-1)^2.
\end{eqnarray}
Examining the behaviour of Kretschmann scalar in the vicinity of horizon we can show that it takes finite value and it means that there is an ordinary coordinate singularity as it should be for a black hole. It was noted above that when $r\rightarrow\infty$ the metric function $U(r)$ behaves as  (\ref{metr_inf_any}), thus Kretschmann scalar takes the form:
\begin{equation}\label{Kr_sc_infty}
\nonumber R_{\mu\nu\k\l}R^{\mu\nu\k\l}\sim \frac{8(n+1)\al^2}{n(n-1)^2\e^2}.
\end{equation}
Finally, when $r\rightarrow 0$ the Kretschmann scalar has very similar dependence of coordinate $r$, for power-law and Born-Infeld cases, except $n\neq 3,4$ and $\ve\neq 0$ for Born-Infeld  case. Namely, we can write:
\begin{eqnarray}\label{Kr_scal_sing}
\nonumber R_{\mu\nu\k\l}R^{\mu\nu\k\l}\sim \frac{1}{r^4}.
\end{eqnarray}
If $n=4$, in the dependence (\ref{Kr_scal_sing}) we have additional logarithmically divergent factor, while for $n=3$ the scalar behaves as $\sim 1/r^6$, this can be explained by the dominance of other terms at small distances in this case. To sum it up, we see that for both types of field and all geometries Kretschmann scalar is divergent at the origin  and it means that we have true physical singularity at this point.
\section{Thermodynamics of the black holes}
\begin{figure}
\centerline{\includegraphics[scale=0.3,clip]{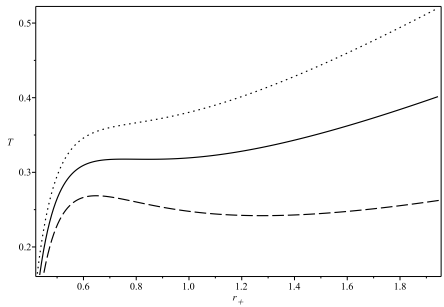}}
\caption{Black hole temperature for PMI case}\label{temp_graph}
\end{figure}

In this  section we give short sketch of black hole thermodynamics for the obtained solutions in the theory with nonminimal coupling. To examine thermodynamics we depart from black hole's temperature which can be calculated using standard approach of General Relativity, so it can be written is as follows:
\begin{eqnarray}\label{temperature}
T=\frac{\kappa}{2\pi}=\frac{1}{4\pi}\frac{U\rq{}(r_+)}{\sqrt{U(r_+)W(r_+)}}.  
\end{eqnarray}
Using the written above relation it is easy to find evident analytical expressions for the temperature, but we do not write these final forms here because of the complicated forms we arrive at, instead we demonstrate the behaviour of the temperature graphically. Namely, on the Fig.[\ref{temp_graph}] the temperature as a function of horizon radius $r_+$ is shown. This figure demonstrates that for large horizon radii the temperature goes up almost linearly, it is caused by the dominance of AdS-terms in this case.  For small $r_+$ the temperature decreases, but for some intermediate values of $r_+$ its behaviour is nonmonotonous, this fact might give some hints about criticality for such system, but we do not investigate this issue here.

To examine thermodynamics of the black holes it is necessary to obtain the expression for the entropy and write the first law of black hole's thermodynamics. Since we consider the theory with nonminimal coupling the expression for the entropy might take some other form than it is in the standard General Relativity. To obtain the expression for the entropy we use well-established Wald's approach \cite{Wald_PRD93}, which can be treated as a variant of Noether method which allows to find conserved quantities. According to the approach to obtain conserved quantities we start from examination of the surface terms which appear when one performs variation of the action integral (\ref{action}).  We do not go deeply into the details of corresponding calculations, but show the final relation which is of the crucial importance here, namely it takes the form:
\begin{equation}\label{wald_eq}
\delta {\cal H}_{\infty}=\delta {\cal H}_{+}
\end{equation} 
and here $\delta {\cal H}$ denotes the variation of the gravitational Hamiltonian and it can be written as follows:
\begin{equation}\label{var_h}
\delta {\cal H}=\delta\int_{c}J_{(n)}-\int_{c}d\left(i_{\xi}\Theta_{(n)}\right)
=\int_{\Sigma^{n-1}}\delta Q_{(n-1)}-i_{\xi}\Theta_{(n)},
\end{equation} 
and here $Q_{(n-1)}$ and $\Theta_{(n)}$ are some forms which are derived from the boundary term in the variation of the action. Omitting the details of calculations we can write the variation of the Hamiltonian at the infinity, for both types of field (power-law and Born-Infeld) we obtain:
\begin{equation}\label{var_H_inf}
\delta{\cal H}_{\infty}=\delta M-\Phi_{q}\delta Q,
\end{equation}
where we have:
\begin{equation}
M=\frac{(n-1)\omega_{n-1}}{16\pi}\mu,
\end{equation} 
is supposed to be the mass of the black holes and 
\begin{equation}\label{total_charge}
Q=\frac{\omega_{n-1}}{4\pi}2^{p-1}q^{2p-1} \quad (PMI), \quad\quad Q=\frac{\omega_{n-1}}{4\pi}q \quad (BI)
\end{equation}
are the total electric charges of two types of black holes respectively. We point out here that the total charges (\ref{total_charge}) can be calculated independently using Gauss law. At the horizon we obtain:
\begin{equation}\label{TD_diff}
\delta {\cal H}_{+}=\frac{(n-1)\omega_{n-1}}{16\pi}U\rq{}(r_+)r^{n-2}_{+}\delta r_{+}=\left(1+\frac{\e}{4}\frac{(\vp\rq{})^2}{W}\Big|_{r_+}\right)T\delta\left(\frac{{\cal A}}{4}\right).
\end{equation}
where ${\cal A}=\omega_{n-1}r^{n-1}_+$ is the horizon area of the black hole. Finally, having used the relation (\ref{wald_eq}) we can write:
\begin{equation}\label{first_law}
\delta M=\left(1+\frac{\e}{4}\frac{(\vp\rq{})^2}{W}\Big|_{r_+}\right)T\delta\left(\frac{{\cal A}}{4}\right)+\Phi_q\delta Q.
\end{equation}
 The written above equation is the first law of black hole thermodynamics, but there are some differences with the first law in the framework of General Relativity, where entropy can be easily introduced as a quarter of horizon area, here we cannot introduce the entropy in the same way unless the temperature is redefined. The relation for the temperature is well-established, so relation for entropy should be modified and to obtain the modified relation for entropy one should introduce additional ``scalar charges'' [12,14]. We introduce entropy as follows:
\begin{equation}\label{entropy}
S=\left(1+\frac{\e}{4}\frac{(\vp')^2}{W}\Big|_{r_+}\right)\frac{{\cal A}}{4},
\end{equation}
and ``scalar charge'' and corresponding to it ``scalar potential'' can be represented in the form:
\begin{equation}\label{sc_pot}
Q^{+}_{\vp}=\omega_{n-1}\sqrt{1+\frac{\e}{4}\frac{(\vp')^2}{W}\Big|_{r_+}}, \quad \Phi^{+}_{\vp}=-\frac{{\cal A}T}{2\omega_{n-1}}\sqrt{1+\frac{\e}{4}\frac{(\vp')^2}{W}\Big|_{r_+}}.
\end{equation}
Having defined entropy, ``scalar charge'' and corresponding potential we can represent the first law of the black hole in the following form:
\begin{equation}\label{first_law_new}
\delta M=T\delta S+\Phi^{+}_{\vp}\delta Q^{+}_{\vp}+\Phi_{q}\delta Q.
\end{equation}
The relation for entropy (\ref{entropy}) allowed us to develop consistent thermodynamics, and to obtain modified form for the first law (\ref{first_law_new}). But there is an obstacle against introduced form for the ``scalar charge'', namely the action (1) is shift-symmetric $\vp\rightarrow\vp+C$ and it means that some conserved charge which reflects this symmetry should exist, but in the black hole solutions we have derived there is no ``charge'' which corresponds to this symmetry. This fact might be explained by a specific condition imposed on the component of Einstein tensor $G_{rr}$ while integrating the equation for the scalar field. Nevertheless the first law of thermodynamics and corresponding values for the black hole's mass, temperature and entropy are consistent, the possible doubts regarding the definition of entropy and ``scalar charge'' can be removed by some independent way of derivation of thermodynamic values, for instance applying Euclidean technique, but it will be performed elsewhere.

\section{Conclusions}
In this work topological static black hole solutions in the theory with nonminimal derivative coupling between scalar field  and gravity and with additional nonlinear electromagnetic field minimally coupled to gravity are obtained.  Detailed analysis of the metric function $U(r)$ shows that its behaviour at infinity is of AdS-type, whereas at the origin all the solutions have the point of physical singularity, what is typical for black holes.

We also give brief sketch of some aspect of black hole thermodynamics. Firstly, temperature of black holes was calculated and even short  glance on the dependence of the temperature shows that some critical behaviour might take place for the black holes. To derive the first law of black hole thermodynamics we have used Wald's procedure which is applicable to quite general diffeomorphism-invariant theories. In contrast with the standard case we are not able to define the entropy of the black hole as a quarter of the horizon area, because an additional factor in the first term of (\ref{first_law}). There are two simplest ways to define the entropy, namely to keep the standard relation from General Relativity, but modify the relation for the temperature and the other way is to define a new relation for the entropy, but the price we have to pay here is to introduce additional ``charges'' as it was performed in \cite{Feng_JHEP15,Stetsko_PRD19}. Here we follow the second way, namely we introduce relation for entropy (\ref{entropy}), define ``scalar charge" together with corresponding ``scalar potential" and finally derive new relation for the first law (\ref{first_law_new}), all these relations are consistent, but as we noted at the end of the previous section there are some drawbacks related to definition of the ``scalar charge". This vagueness as it was also noted above can be removed when one uses an independent way to
obtain thermodynamic values.

\section{Acknowledgements}
This work was partly supported by Project FF-83F (No. 0119U002203) from the Ministry of Education and Science of Ukraine.


\begin{thebibliography}{99}
\bibitem{Will_LRR2014} C.~M.~Will, Liv.~Rev.~Rel. {\bf 17}, 4 (2014).
\bibitem{Berti_CQG2015} E.~Berti, E.~Barausse, V.~Cardoso, {\it et al.} Class.~Quant.~Grav. {\bf 32}, 243001 (2015). 
\bibitem{Clifton_PhysRept2012} T.~Clifton, P.~G.~Ferreira, A.~Padilla, C.~Skordis, Phys.~Rept. {\bf 513}, 1 (2012).
\bibitem{Heisenberg_PhRept2019} L.~Heisenberg, Phys.~Rept. {\bf 796}, 1, (2019).
\bibitem{Horndeski_IJTP74} G.~W.~Horndeski, Int.~Journ.~Theor.~Phys. {\bf 10}, 363, (1974).
\bibitem{Deffayet_PRD09} C.~Deffayet, S.~Deser, G.~Esposito-Farese, Phys.~Rev.~D {\bf 80}, 064015 (2009).
\bibitem{Sushkov_PRD09} S.~V.~Sushkov, Phys.~Rev.~D {\bf 80}, 103505 (2009).
\bibitem{Saridakis_PRD10} E.~N.~Saridakis, S.~V.~Sushkov, Phys.~Rev.~D {\bf 81}, 083510 (2010).
\bibitem{Germani_PRL10} C.~Germani, A.~Kehagias, Phys.~Rev.~Lett. {\bf 105}, 011302 (2010); C.~Germani, A.~Kehagias, Phys.~Rev.~Lett. {\bf 106}, 161302 (2011).
\bibitem{Rinaldi_PRD12} M.~Rinaldi, Phys.~Rev.~D {\bf 86}, 084048 (2012).
\bibitem{Minamitsuji_PRD14} M.~Minamitsuji, Phys.~Rev.~D {\bf 89}, 064017, (2014); E.~Babichev, C.~Charmousis, JHEP {\bf 08}, 106 (2014); A.~Anabalon, A.~Cisterna, J.~Oliva, Phys.~Rev.~D {\bf 89}, 084050 (2014); T.~Kobayashi, N.~Tanahashi, Prog.~Theor.~Exp.~Phys. {\bf 2014}, 073E02 (2014); M.~Bravo-Gaete, M.~Hassaine, Phys.~Rev.~D {\bf 90}, 024008 (2014). T.~P.~Sotiriou, S.-Y.~Zhu, Phys.~Rev.~Lett. {\bf 112}, 251102 (2014). A.~Maselli, H.~O.~Silva, M.~Minamitsuji, E.~Berti, Phys.~Rev.~D~{\bf 92}, 104049 (2015). G.~Giribet, M.~Tsoukalas, Phys.~Rev.~D {\bf 92}, 064027 (2015).
\bibitem{Feng_JHEP15} X.-H.~Feng, H.-S.~Liu, H.~Lu, C.~N.~Pope, JHEP {\bf 1511}, 176 (2015). X.-H.~Feng, H.-S.~Liu, H.~Lu, C.~N.~Pope, Phys. Rev. D {\bf 93} 044030 (2016). 
\bibitem{Stetsko_arx18} M.~M.~Stetsko, arXiv:1811.05030;
\bibitem{Stetsko_PRD19} M.~M.~Stetsko, Phys.~Rev.~D {\bf 99}, 044028 (2019).
\bibitem{Wald_PRD93} R.~M.~Wald, Phys. Rev. D {\bf 48}, 3427 (1993); V.~Iyer, R.~M.~Wald, Phys. Rev. D {\bf 50}, 846 (1994).
\end{thebibliography}
\end{document}